\documentclass{sigchi-ext}
\usepackage[T1]{fontenc}
\usepackage{textcomp}
\usepackage[scaled=.92]{helvet} 
\usepackage{graphicx} 
\usepackage{balance}  
\usepackage{booktabs} 
\usepackage{ccicons}  
\usepackage{ragged2e} 

\def\plaintitle{Human-centred Explainability and Explainable AI} 
  
\def\emptyauthor{}
\def\plainkeywords{Human-centered Explainability; xAI; Algebraic Machine Learning }

\title{Combining Human-centred Explainability and\\ Explainable AI}

\numberofauthors{2}
\author{%
  \alignauthor{%
    \textbf{Janin Koch}\\
    \affaddr{Université Paris-Saclay, CNRS,Inria,LISN} \\
    \affaddr{Paris, France} \\
    \email{Janin.Koch@inria.fr} }
    \alignauthor{%
    \textbf{Vitor Fortes Rey}\\
    \affaddr{German Research Center for Artificial
Intelligence (DFKI)}\\
    \affaddr{University of Kaiserslautern, Germany}\\
    \email{fortes@dfki.uni-kl.de} } }

\definecolor{linkColor}{RGB}{6,125,233}
\hypersetup{%
  pdftitle={\plaintitle},
  pdfauthor={\emptyauthor},
  pdfkeywords={\plainkeywords},
  bookmarksnumbered,
  pdfstartview={FitH},
  colorlinks,
  citecolor=black,
  filecolor=black,
  linkcolor=black,
  urlcolor=linkColor,
  breaklinks=true,
}

\begin{document}

\CopyrightYear{2020}
\setcopyright{rightsretained}
\conferenceinfo{CHI'20,}{April  25--30, 2020, Honolulu, HI, USA}
\isbn{978-1-4503-6819-3/20/04}
\doi{https://doi.org/10.1145/3334480.XXXXXXX}
\copyrightinfo{\acmcopyright}

\maketitle

\RaggedRight{} 

\begin{abstract}
  This position paper looks at differences between the current understandings of human-centered explainability and explainability AI. We discuss current ideas in both fields, as well as the differences and opportunities we discovered. As an example of combining both, we will present preliminary work on a new algebraic machine learning approach. We are excited to continue discussing design opportunities for human-centered explainability (HCx) and xAI with the broader HCxAI community.
\end{abstract}




\section{Introduction}
Human-AI collaboration, communication, and interaction research is critical for advances in enhancing, and expanding human capabilities with technology.
Roschelle et al. define collaboration as `a coordinated, synchronous activity that is the result of a continued attempt to construct and maintain a shared conception of a problem'~\cite{roschelle1995construction}. Empirical evidence show that such collaborative learning enhances the cognitive capabilities of the people involved, allowing them as a team to reach a level of cognitive performance that exceeds the sum of the individuals'~\cite{baker2015collaboration}.

This collaborative potential has attracted the interest of the HCI and AI research communities, especially in the area of advancing human-AI collaborative approaches.
These collaborative effects, however, are based on interaction and common understanding, as described by Rochelle et al. definition of as 'continued attempt to construct and maintain a shared conception of a problem'~\cite{roschelle1995construction}.
As a result, in order to investigate these aspects, we took a step back and focused on how to form a common understanding between humans and machines, from both a human-centric and an AI-centric perspective and how to build explainable approaches to support it.

In this position paper we will briefly outline our understanding of human-centered and xAI and the differences and opportunities we see. We will then present our preliminary work on a new algebraic machine approach as an example to combined both.

\section{Explainable AI (xAI)}
The objective of an explainable AI system is to offer explanations for its behavior so that humans may better understand it.
Gunning et al. offer the following key factors to improve the design of these systems~\cite{gunning2019xai}, which we believe provides a good insight into the perspective of xAI.
An xAI system should:
\begin{itemize}[noitemsep]
    \item describe its capabilities and understandings within the current context;
    \item explain what it has done, what it is doing now, and what will happen next; 
    \item disclose the underlying information it is acting upon~\cite{bellotti2001intelligibility}
\end{itemize}

These factors are critical for gaining a better understanding of the system's internal mechanisms. 'Black-box' algorithms, in particular Deep Learning approaches, commonly struggle to provide such information. However, there are numerous approaches that aim to live up to these key factors.
Angelov and colleagues outline a number of such approaches in a review on explainable artificial intelligence~\cite{angelov2021explainable}, and then provide an xAI taxonomy based on their findings. However, these approaches continue to focus primarily on domain experts in order to better understand the systems' reasoning rather than how the AI output can be interpreted in the current context of use for a user.

\section{Human-Centered Explainability (HCx)}
Research in human-centred explainability aims to contextualize intelligent system contributions in the \textit{user's ongoing task and understanding}. The human-centered artificial intelligence (HCAI) framework~\cite{shneiderman2020human} clarifies that human-centered design of AI and AI interaction should: 

\begin{itemize}[noitemsep]
\item  Design for high levels of human control as well as high levels of computer automation in order to improve human performance.
\item  Recognize the situations in which full human control or full computer control is required.
\item  Avoid the dangers of having too much human control or too much computer control.
\end{itemize}
Shneiderman et al. outlined in their overview, how these factors are necessary to create Reliable, Safe and Trustworthy intelligent systems, and highlight especially the role of explainability and controllability in this context~\cite{shneiderman2020human}. 

There is already a body of work in HCI that seeks to implement these guidelines. Research, particularly in the context of creativity support tools and co-creative systems, emphasizes the need for more elaborated and contextualized explanations of contributions.
Zhu et al. provide a detailed overview of explainable AI for designers and co-creative practices~\cite{zhu2018explainable} and even propose to expand research in this area in a new research area called eXplainable AI for designers (XAID).
A good example is MayAI~\cite{koch2019may}, an AI system that develops ideas together with a designer. Its outstanding features are the AI’s ability to adapt to evolving objectives, and to make its own decisions to explore or exploit ideas. It contains a separately developed explainable engine to explain the systems contribution to the ideation process based on the relevance for the designer. It does consider rather semantic and visually perceivable information for such explanations, instead of explaining learned relations. This was highlighted as very valuable in an realistic ideation study with the system. 

While creativity is a rather specific use case, we think that the lessons learned from such open-ended tasks are useful for designing general human-AI interaction. While the focus and approach differ, we believe that xAI approaches, i.e. why does a system decide in a certain way, and human-centered explainability, i.e. how is that relevant for the user, can be combined.
Though this necessitates the development of machine learning (ML) approaches with a focus on human-centered interpretability, it would not only allow for a better understanding of the algorithms' inner mechanisms, but it would also open up opportunities to map system output to contextual and relatable explanations.

\section{Contrasting HCx and xAI}
To summarize, human-centered explainability (HCx) aims to contextualize intelligent system contributions in the \textit{user's ongoing task and understanding}. Common explainable AI (XAI) approaches, in contrast, aim to explain the underlying \textit{computational decision making} to users. 

The difference between the two lies in the \textbf{context} of the explanation: xAI refers to the way the system works and reaches decisions, while human-centered explainability seeks to fulfill a need for explanations as to why a given result is relevant to the current task. 
Suggestion engines, such as those found in online stores, are a good example. An xAI approach could attempt to explain which parameters were taken into account to result in the final suggestion, such as the technical equipment used for the search, other websites visited, or previous purchases. A more human-centered explainability could outline how the suggestion might fit \textit{into} current searches, or how it might interact with other products purchased. It could further include overviews of the benefits of the suggestion over other items currently being examined. This allows users to better comprehend the suggestions in the context of their own goals, and it opens up opportunities to steer or lead the human-AI interaction to a more desired outcome.

The distinction between HCx and xAI thus differs in method and goal of explainability. Collecting parameters of a \textit{user's perception} of a system's contribution to a task (HCx) can help to make AI contributions more understandable and relevant to end users so that they can attain high levels of control. This could result in more trustable intelligent systems that are more embedded in the context of use.
Focusing on expressing parameters of how the system \textit{"works"} to help professionals better understand the algorithm, on the other hand, can help identify biases and errors, as well as improve model quality.
While this information is useful for understanding the algorithms, it is often less relevant to the end user's intent, which may lead to information fatigue and a decrease in trustworthiness. As a result, we argue that the goal of explainable should be considered early on, in order to determine which parameters must be collected to facilitate it. Combining the two methods entails gathering and combining system and usage context data that can be used to develop a more comprehensive understanding of AI contributions.

\section{Combining HCx and xAI using AML-interpretability}
As a final part of our contribution, we'd like to highlight our most recent work on human-centered explainability in combination with the concept of xAI. For this approach, we take advantage of the interpretability capabilities of Algebraic machine learning (AML) models.
AML-models are algebraic representations in a mathematical model~\cite{17martin2018algebraic}. The foundation of algebraic machine learning (AML) is a set of essential algorithms that can learn from data. Individual AML applications for specialized analyses need the use of the AML description language (AML-DL) to construct embeddings with human-defined limitations. During training, the core learning algorithms continue to generate atoms until the model learns to distinguish between classes while adhering to the AML-DL restrictions established by the user.

In comparison to machine learning models such as deep neural networks, which are recognized as being less suitable for interpretability and explainability in the context of explainable AI (xAI)~\cite{18rudin2019stop}, AML models have three potential advantages:
\begin{itemize}[noitemsep]
    \item AML-models have a fundamentally simple structure. They comprise only three layers: inputs – atoms – outputs. An atom can be linked to one or more inputs and is said to be present in an example if at least one of those inputs is present. AML-models expand as they learn patterns from training data, though the simple three-layer structure persists.
    
    \item The core AML algorithms check that the human-defined constraints are maintained in the subsequent binary relationships between inputs and outputs that are learned as new atoms are generated during training. The overall integrity of the AML-DL model is maintained however large the number of atoms grows to be.
    
    \item AML-DL rules can be traced back to human-defined policies, for example, those that  provide the basis for human-defined work procedures. That is, humans can interpret how the model uses inputs to generate outputs~\cite{19lundberg2017unified} and how to control it in the current context.
\end{itemize}

While all three aspects of AML are crucial for explainability, the last highlights this potential. The final atom space representation can then be converted into a decision tree using the CART algorithm \cite{loh2011classification}, limiting its height so that said tree is easier to interpret. 
\begin{figure}[t]
    \centering
    \includegraphics[width=0.3\textwidth]{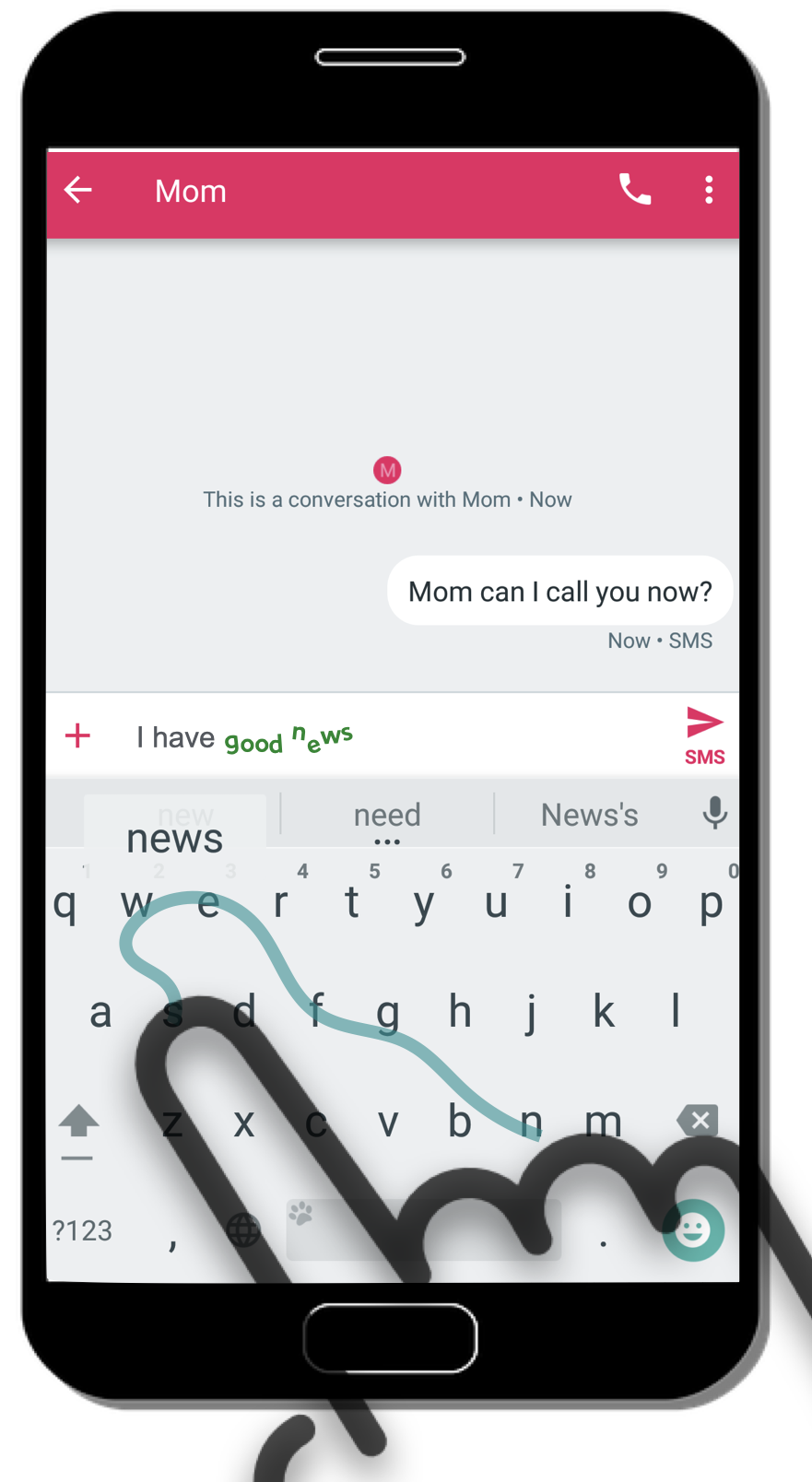}
    \caption{Interface example for expressive gesture typing}~\label{fig:one}
\end{figure}

\subsection{Gesture typing application}
We tested this potential of interpretability on a use case of gesture typing. In this example the raw position data were streamed to the AML-model, which in return provides values of the features used to predict the recognized class, as well as the prediction itself and its confidence level for this recognition. Our mobile application then maps these features to expressive forms of e.g. type and style. An example of such human-AI interaction is shown in Figure \ref{fig:one}. 

The data can also be used to visualize the xAI features used to classify the data, such as gesture speed or acceleration. The atom structure also allows us to generalize the task, situation, and context of gesturing. This allows us to provide more human-centered explainability (HCx) of how user actions directly affect the font and color selection. We can provide guidance on how to control this interpretation given the current situation by displaying alternative gesture spaces for the user's intended goals based on this more general understanding of interpretation.
This project is currently in progress, and it will be published alongside an in-depth assessment of the AML-model approach's explainability potential.

\section{Discussion}
We believe there is a significant opportunity for future research into merging explainability techniques that focus on better understanding how AI algorithms work and how their outcomes can be more relevant, controlled, and trustworthy for end users.
This requires that designers of interactive machine learning systems clarify the goal of explainability early on and how it can be included into the system's inner workings. We showed how the intrinsic structure of an ML approach might integrate such efforts, but we see a lot of possibilities for using and combining other existing methods as well.
We hope that our preliminary work will encourage the development of new ML approaches that are designed with human-interpretability in mind.

\section{Acknowledgements}
We thank all our collaborators on the ALMA project. This work was funded by the European Research Council (ERC) under the 'ALMA: Human Centric Algebraic Machine Learning' Project
(grant agreement No. 952091).

\balance{} 

\bibliographystyle{references}
\bibliography{sample}


\begin{thebibliography}{00}


\ifx \showCODEN    \undefined \def \showCODEN     #1{\unskip}     \fi
\ifx \showDOI      \undefined \def \showDOI       #1{{\tt DOI:}\penalty0{#1}\ }
  \fi
\ifx \showISBNx    \undefined \def \showISBNx     #1{\unskip}     \fi
\ifx \showISBNxiii \undefined \def \showISBNxiii  #1{\unskip}     \fi
\ifx \showISSN     \undefined \def \showISSN      #1{\unskip}     \fi
\ifx \showLCCN     \undefined \def \showLCCN      #1{\unskip}     \fi
\ifx \shownote     \undefined \def \shownote      #1{#1}          \fi
\ifx \showarticletitle \undefined \def \showarticletitle #1{#1}   \fi
\ifx \showURL      \undefined \def \showURL       #1{#1}          \fi

\bibitem{angelov2021explainable}
{Plamen~P Angelov}, {Eduardo~A Soares}, {Richard Jiang}, {Nicholas~I Arnold},
  {and} {Peter~M Atkinson}. 2021.
\newblock \showarticletitle{Explainable artificial intelligence: an analytical
  review}.
\newblock {\em Wiley Interdisciplinary Reviews: Data Mining and Knowledge
  Discovery\/} {11}, 5 (2021), e1424.
\newblock


\bibitem{baker2015collaboration}
{Michael~J Baker}. 2015.
\newblock \showarticletitle{Collaboration in collaborative learning}.
\newblock {\em Interaction Studies\/} {16}, 3 (2015), 451--473.
\newblock


\bibitem{bellotti2001intelligibility}
{Victoria Bellotti} {and} {Keith Edwards}. 2001.
\newblock \showarticletitle{Intelligibility and accountability: human
  considerations in context-aware systems}.
\newblock {\em Human--Computer Interaction\/} {16}, 2-4 (2001), 193--212.
\newblock


\bibitem{gunning2019xai}
{David Gunning}, {Mark Stefik}, {Jaesik Choi}, {Timothy Miller}, {Simone
  Stumpf}, {and} {Guang-Zhong Yang}. 2019.
\newblock \showarticletitle{XAI—Explainable artificial intelligence}.
\newblock {\em Science Robotics\/} {4}, 37 (2019), eaay7120.
\newblock


\bibitem{koch2019may}
{Janin Koch}, {Andr{\'e}s Lucero}, {Lena Hegemann}, {and} {Antti Oulasvirta}.
  2019.
\newblock \showarticletitle{May AI? Design ideation with cooperative contextual
  bandits}. In {\em Proceedings of the 2019 CHI Conference on Human Factors in
  Computing Systems}. 1--12.
\newblock


\bibitem{loh2011classification}
{Wei-Yin Loh}. 2011.
\newblock \showarticletitle{Classification and regression trees}.
\newblock {\em Wiley interdisciplinary reviews: data mining and knowledge
  discovery\/} {1}, 1 (2011), 14--23.
\newblock


\bibitem{19lundberg2017unified}
{Scott~M Lundberg} {and} {Su-In Lee}. 2017.
\newblock \showarticletitle{A unified approach to interpreting model
  predictions}.
\newblock {\em Advances in neural information processing systems\/}  {30}
  (2017).
\newblock


\bibitem{17martin2018algebraic}
{Fernando Martin-Maroto} {and} {Gonzalo~G de Polavieja}. 2018.
\newblock \showarticletitle{Algebraic Machine Learning}.
\newblock {\em arXiv preprint arXiv:1803.05252\/} (2018).
\newblock


\bibitem{roschelle1995construction}
{Jeremy Roschelle}, {Stephanie~D Teasley}, {and} {others}. 1995.
\newblock \showarticletitle{The construction of shared knowledge in
  collaborative problem solving}. In {\em Computer-supported collaborative
  learning}, Vol. 128. 69--197.
\newblock


\bibitem{18rudin2019stop}
{Cynthia Rudin}. 2019.
\newblock \showarticletitle{Stop explaining black box machine learning models
  for high stakes decisions and use interpretable models instead}.
\newblock {\em Nature Machine Intelligence\/} {1}, 5 (2019), 206--215.
\newblock


\bibitem{shneiderman2020human}
{Ben Shneiderman}. 2020.
\newblock \showarticletitle{Human-centered artificial intelligence: Reliable,
  safe \& trustworthy}.
\newblock {\em International Journal of Human--Computer Interaction\/} {36}, 6
  (2020), 495--504.
\newblock


\bibitem{zhu2018explainable}
{Jichen Zhu}, {Antonios Liapis}, {Sebastian Risi}, {Rafael Bidarra}, {and}
  {G~Michael Youngblood}. 2018.
\newblock \showarticletitle{Explainable AI for designers: A human-centered
  perspective on mixed-initiative co-creation}. In {\em 2018 IEEE Conference on
  Computational Intelligence and Games (CIG)}. IEEE, 1--8.
\newblock


\end{thebibliography}

\end{document}